# Transcriptional Regulation by the Numbers 1: Models

Lacramioara Bintu[1], Nicolas E. Buchler[2], Hernan G. Garcia[3], Ulrich Gerland[4], Terence Hwa[5], Jané Kondev[1], and Rob Phillips[6]

[1]*Physics Department, Brandeis University, Waltham, MA 02454*

[2]*Center for Studies in Physics and Biology, The Rockefeller University, New York, NY 10021*

[3]*Department of Physics, California Institute of Technology, Pasadena, CA 91125*

[4]*Physics Department and CENS, Ludwig-Maximilians University, Munich, Germany*

[5]*Physics Department and Center for Theoretical Biological Physics, University of California at San Diego, La Jolla, CA 92093-0374*

[6]*Division of Engineering and Applied Science and Kavli Nanoscience Institute, California Institute of Technology, Pasadena, CA 91125*

Abstract

The study of gene regulation and expression is often discussed in quantitative terms. In particular, the expression of genes is regularly characterized with respect to how much, how fast, when and where. Whether discussing the level of gene expression in a bacterium or its precise location within a developing embryo, the natural language for these experiments is that of numbers. Such quantitative data demands quantitative models. We review a class of models ("thermodynamic models") which exploit statistical mechanics to compute the probability that RNA polymerase is at the appropriate promoter. This provides a mathematically precise elaboration of the idea that activators are agents of recruitment which increase the probability that RNA polymerase will be found at the promoter of interest. We discuss a framework which describes the interactions of repressors, activators, helper molecules and RNA polymerase using the concept of effective concentrations, expressed in terms of a function we call the "regulation factor". This analysis culminates in an expression for the probability of RNA polymerase binding at the promoter of interest as a function of the number of regulatory proteins in the cell. In a companion paper [1], these ideas are applied to several case studies which illustrate the use of the general formalism.

## 1. Introduction

The biological literature on the regulation and expression of genes is with increasing frequency couched in the language of numbers. Four key ways in which gene expression is characterized quantitatively is through measurement of: i) the level of expression relative to some reference value, ii) how fast a given gene is expressed after induction, iii) the precise relative timing of expression of different genes, and iv) the spatial location of expression. Our aim in this first section is to revisit particular examples of such measurements in the bacterial setting, which provide the motivation for the models that form the main substance of this and the companion article. Note that through much of the article we call attention to particular revealing case studies and do not give a thorough coverage of the literature.

*How much, when and where?* One particularly well characterized class of examples of the level of gene expression which will serve as the centerpiece of this and the companion article are those associated with bacterial metabolism and the infection of bacteria by phage [2\*\*, 3\*\*]. In the classic case of the *lac* operon a number of beautiful measurements have been performed which characterize the extent to which the genes are repressed as a function of the strength of the operators, their spacing, and on the number of repressor molecules [4, 5, 6]. Similar measurements have been performed for other genes implicated in bacterial metabolism as well as for those tied to the decision between the lytic and lysogenic pathways after infection of *E. coli* by phage lambda [7, 8, 9, 10, 11]. A second way in which measurements on the regulatory status of a given system are quantitative is to measure *when* genes of interest are being expressed. The list of examples is both long and inspiring and several representative case studies can be found in [12, 13, 14]. A third way in which an increasingly quantitative picture of gene expression is emerging is based on the ability to make precise statements about the spatial location of the expression of different genes. Here too, the number of different examples that can be mustered to prove the general point is staggering [15\*\*, 16\*, 17]. The key point of these examples is to note the growing pressure head of quantitative *in vivo* data, which calls for more than a cartoon-level description of expression.

The physicochemical modeling of the type of quantitative data described above is still in its infancy. One class of models, which will serve as the basis of the present article, are the so-called "thermodynamic models" [18, 19, 20]. The conceptual basis of this class of models is the idea that the expression level of the gene of interest can be deduced by examining the equilibrium probabilities that the DNA associated with that gene is occupied by various molecules, such as RNA polymerase and a battery of transcription factors, such as repressors and activators. There is a longstanding tradition of using these ideas, with particularly important examples associated with the famed examples of the *lac* operon and phage lambda [18, 21, 22, 23, 24, 25]. Note also that the thermodynamic models can serve as input to more general chemical kinetic models.

The key aim of this and the accompanying article [1] is to show how the "thermodynamic models" yield a general conceptual picture of regulation using what we will call the "regulation factor". Such arguments are useful because they permit a direct confrontation with quantitative experiments, like those discussed above. The purpose of models is not just to "fit the data" (though such fits can reveal which mechanisms are operative), but also to provide a conceptual scheme for understanding measurements, and more importantly, for suggesting new experiments. It is also worth noting that when such models fall short, it is an opportunity to find out why and learn something new.

The logic of this and the following article is as follows. The first article is to a large extent pedagogical and aims to show how a microscopic picture of the various states of the gene of interest can be mathematicized using statistical mechanics. The companion article is built around the analysis of real case studies in bacterial transcription and centers concretely on how the activity of a given promoter is altered (the *fold-change* in promoter activity) by the presence of transcription factors.

**2. "Thermodynamic Models" of Gene Regulation: The Regulation Factor**

The fundamental tenet of the "thermodynamic models" for gene regulation is that we can replace the difficult question of computing the level of gene expression, as measured by the concentration of gene product ([*protein*]), with the more tractable question of the probability ($p_{bound}$) that RNA polymerase occupies the promoter of interest. More precisely, these models are founded on the idea that the instantaneous disposition of the gene of interest can be obtained from the probability that various molecules (RNAP, activators, repressors, inducers) are bound to their relevant targets.

Such models are based on a variety of different assumptions, all of which can and should be evaluated critically. Perhaps the most glaring assumption is that of equilibrium itself. This assumption can be examined quantitatively on the basis of the relative rates of transcription factor (TF) binding, RNA polymerase binding, open complex formation, transcript formation and translation itself. For example, if the rate for open complex formation is much smaller than the rates for RNA polymerase getting on and off the promoter, then the probability for binding the polymerase on the promoter will be given by its equilibrium value. A second key assumption of this class of models is the idea that the probability of promoter occupancy by RNA polymerase is simply proportional to the level of expression of a given gene. The difficulty lies in the fact that there are a number of different mechanisms that can intervene between RNAP binding and the existence of a functional gene product. Despite these caveats, we argue that this class of models is both instructive and predictive, and in those cases where they are found wanting, it is an opportunity to learn something.

Our point of departure is an analysis of the probability that RNAP will be bound at the promoter of interest in the absence of any activators or repressors. This will be followed by cases of increasing complexity that involve batteries of transcription factors. Note also that although our preliminary discussion is focused on the statistical mechanics of polymerase binding, the framework is the same for generic protein-DNA and protein-protein interactions. For the purposes of paper 1, we make the simplifying assumption that the key molecular players (RNAP and repressors) are bound either nonspecifically or specifically to the DNA itself. This question has been addressed in the context of the *lac* operon [21] and the λ switch [26]. Stated differently, as a simplification, we will ignore the contribution of "free" polymerase in the cytoplasm, as well as those RNAP molecules that are engaged in transcription on other promoters. Relaxing this assumption has no effect on the framework developed below. Hence, to evaluate the probability of promoter occupancy in this simple model, the reservoir of RNA polymerases will be the nonspecifically bound molecules as shown in fig.1(a).

To evaluate the probability of polymerase binding ($p_{bound}$), we need to sum the Boltzmann weights over all possible states of *P* polymerase molecules on DNA [27**, 28**]. We note that *P* should really be thought of as an effective number of RNAP molecules available for binding to the promoter. Estimating this number *in vivo* is fraught with difficulty as many RNAPs are engaged in transcription at any given time, and as such are not available for binding. Fortunately, when computing the fold-change in activity of a given promoter, as we do in paper 2 [1], for all cases of interest, namely when $p_{bound}$ in the absence of activators is very small, *P* drops out of the problem.

The calculation of $p_{bound}$ goes as follows. We consider *P* RNAP molecules distributed amongst the $N_{NS}$ nonspecific sites which make up the genome itself, and a single promoter. Then we distinguish two classes of outcomes, shown in fig. 1(b): all $P$ RNAP molecules bound nonspecifically, or one RNAP on the promoter and $P$-1 bound nonspecifically. Next, we count the number of different ways that these outcomes can be realized. Once the states have been enumerated, we weight them each according to the Boltzmann law: if $\varepsilon$ is the energy of a state, its statistical weight is $exp(-\varepsilon/k_BT)$. Finally, to compute the probability of promoter occupancy, we construct the ratio of the sum of the weights for the favorable outcome (i.e. promoter occupied) to the sum over all of the weights.

As noted above, in this simple model there are two broad classes of microscopic outcomes: i) those in which all $P$ polymerase molecules are distributed amongst the nonspecific sites and ii) those in which the promoter is occupied and the remaining $P$-1 polymerase molecules are distributed amongst the nonspecific sites. To evaluate the probabilities of these two eventualities, we need to know the number of different ways that each outcome can be realized. The statistical question of how many ways are there to distribute $P$ polymerase molecules amongst $N_{NS}$ nonspecific sites on the DNA is a classic problem in statistics, and the result is $\dfrac{N_{NS}!}{P!(N_{NS}-P)!}$. The overall statistical weight of these states is based not just on how many of them there are, but also on their Boltzmann weights according to

$$\underbrace{Z(P)}_{\text{statistical weight - promoter unoccupied}} = \underbrace{\frac{N_{NS}!}{P!(N_{NS}-P)!}}_{\text{number of arrangements}} \times \underbrace{e^{-P\varepsilon_{pd}^{NS}/k_BT}}_{\text{Boltzmann weight}}, \qquad (1)$$

where $\varepsilon_{pd}^{NS}$ is an energy that represents the average binding energy of RNAP to the genomic background. The correct treatment of the genomic background requires explicit consideration of the distribution of binding energies of RNAP (and TFs) to different sites (both specific and nonspecific) on the DNA. The question of how to treat this problem more generally than the simple minded treatment given here can be found in [29,30]. The *total* statistical weight can now be written as

$$\underbrace{Z_{tot}(P)}_{\text{total statistical weight}} = \underbrace{Z(P)}_{\text{promoter unoccupied}} + \underbrace{Z(P-1)e^{-\varepsilon_{pd}^{S}/k_BT}}_{\text{RNAP on promoter}}, \qquad (2)$$

where $\varepsilon_{pd}^{S}$ is the binding energy for RNAP on the promoter (the $S$ stands for "specific"). The states and corresponding weights, normalized by the weight of the promoter-unoccupied states, $Z(P)$, are shown in fig.1(b).

To find the probability that RNAP is bound to the promoter of interest, we compute

$$p_{bound} = \frac{Z(P-1)e^{-\varepsilon_{pd}^{S}/k_BT}}{Z_{tot}(P)}. \qquad (3)$$

Note that the numerator in this case is the statistical weight of all microscopic states in which the promoter is occupied, while the denominator is the statistical weight of all microscopic states. If we now divide top and bottom by $Z(P-1)e^{-\varepsilon_{pd}^{S}/k_BT}$, and use the

functional form given in eqn.1, the probability of promoter occupancy is given by the simple form

$$p_{bound} = \frac{1}{1 + \frac{N_{NS}}{P} e^{\Delta \varepsilon_{pd}/k_B T}} \quad , \tag{4}$$

where we have introduced the notation $\Delta \varepsilon_{pd} = \varepsilon_{pd}^{S} - \varepsilon_{pd}^{NS}$ [31]. To obtain the last equation we made the simplifying assumption $P \ll N_{NS}$. The results computed above can be depicted in graphical form as shown in fig.1(c) which plots the probability of promoter occupancy as a function of the number of RNAP molecules for two different promoters. For this particular case, we have used several rough estimates (explained in the caption), concerning the binding energies of RNAP molecules to specific and nonspecific sites on the DNA in a typical bacterial cell. One interesting speculation is that the high probability of RNAP occupancy for the T7 promoter, even in the absence of transcription factors, could be related to the infection mechanism of T7 phage [32]. By way of contrast, it is also interesting to note the very low probability of occupancy of the *lac* promoter in this simple model in the absence of activation. We view eqn.4 as characterizing the "basal" transcription rate in this simple model. In light of this result, the key conceptual outcome of the remainder of the paper is the idea that the presence of transcription factors (activators, repressors, etc.) has the effect of altering eqn.4 to the simple form

$$p_{bound} = \frac{1}{1 + \frac{N_{NS}}{P F_{reg}} e^{\Delta \varepsilon_{pd}/k_B T}} \quad , \tag{5}$$

where we introduce the "regulation factor" $F_{reg}$. The regulation factor may be seen as describing an effective increase (for $F_{reg} > 1$), or decrease (for $F_{reg} < 1$), of the number of RNA polymerase molecules that are available to bind the promoter.

To illustrate the idea of the regulation factor concretely, we show how activators recruit [3**] RNA polymerase to the promoter of interest. The recruitment concept is illustrated in cartoon form in fig.2(a), where it is seen that the activator molecule recruits the polymerase through favorable contacts characterized by an adhesive energy $\varepsilon_{ap}$. The point of the cartoon is to show how the various states of occupancy of the promoter and activator binding site may be assigned Boltzmann weights, which can then be used to compute their probabilities.

Once again, the first step in our analysis is to write the total statistical weight. Note that this is obtained by summing the Boltzmann weights of all of the eventualities associated with the activators and polymerase molecules being distributed on the DNA (both nonspecific sites and the promoter). As seen in fig.2(a), there are four classes of outcomes, namely, both the activator site and promoter unoccupied, just the promoter occupied by polymerase, just the activator site occupied by activator and finally, both of the specific sites occupied. This is represented mathematically as

$$Z_{tot}(P, A) = \underbrace{Z(P, A)}_{\text{empty sites}} + \underbrace{Z(P-1, A) e^{-\varepsilon_{pd}^{S}/k_B T}}_{\text{RNAP on promoter}} + \underbrace{Z(P, A-1) e^{-\varepsilon_{ad}^{S}/k_B T}}_{\text{activator on specific site}} \\ + \underbrace{Z(P-1, A-1) e^{-(\varepsilon_{pd}^{S} + \varepsilon_{ad}^{S} + \varepsilon_{pa})/k_B T}}_{\text{RNAP and activator bound specifically}} \quad , \tag{6}$$

where the statistical weight for *P* polymerase molecules and *A* activator molecules distributed among $N_{NS}$ nonspecific sites is given by

$$Z(P,A) = \underbrace{\frac{N_{NS}!}{P!A!(N_{NS}-P-A)!}}_{\text{number of arrangements}} \times \underbrace{e^{-P\varepsilon_{pd}^{NS}/k_BT} e^{-A\varepsilon_{ad}^{NS}/k_BT}}_{\text{weight of each state}} \quad . \tag{7}$$

In fig.2(a) the weights of the four states are normalized by the weight of the empty state $Z(P,A)$. In eqn.7 we use the notation $\varepsilon_{xd}$ to characterize the binding energy of molecule X to DNA, with superscript *S* or *NS* to signify specific or nonspecific binding, respectively, while $\Delta\varepsilon_{xd} = \varepsilon_{xd}^S - \varepsilon_{xd}^{NS}$ is the difference between the two. Note that for the purposes of this simple model we have assumed that the reservoir for the activator molecules is the genomic DNA, though there is strong evidence that in the case of the *lac* operon many of the activators (CRP) are actually in the cytoplasm [38]. By way of contrast, as will be seen in paper 2 [1], in our actual applications of thermodynamic models to real operons, the question of whether the reservoir is nonspecific DNA or the cytoplasm never arises.

As usual, to compute the probability of interest, we construct the ratio of the sum of weights for all those outcomes that are favorable (i.e. polymerase bound to the promoter) to the sum of weights over the total set of outcomes $Z_{tot}(P,A)$. This results in $p_{bound}$ that adopts precisely the form described in eqn.5. The regulation factor, $F_{reg}(A)$, is given by

$$F_{reg}(A) = \frac{1 + \frac{A}{N_{NS}} e^{-\Delta\varepsilon_{ad}/k_BT} e^{-\varepsilon_{ap}/k_BT}}{1 + \frac{A}{N_{NS}} e^{-\Delta\varepsilon_{ad}/k_BT}} \quad , \tag{8}$$

where we have made the additional assumption that $N_{NS} \gg P, A$. Note that in the limit that the adhesive interaction between polymerase and activator goes to zero, the regulation factor itself goes to unity. Further, note that for negative values of this adhesive interaction (i.e. activator and polymerase like to be near each other) the regulation factor is greater than one, which is translated into an apparent increase in the number of polymerase molecules available for binding to the promoter. This claim can be seen more concretely if we define the *fold-change* in promoter activity as the ratio of the probability that RNAP is bound in the presence of transcription factors, to the probability that it is bound in the absence of transcription factors: *fold-change* = $p_{bound}(P, A)/p_{bound}(P, A = 0)$. The *fold-change* is plotted in fig.2(b) for reasonable values of the adhesive interaction $\varepsilon_{ap}$ and the other binding parameters, for the simple model in which the reservoir for CRP is assumed to be nonspecific DNA.

Similar arguments may be made for the action of a repressor molecule. Consider repression by *R* repressor molecules that can bind to an operator (with energy $\varepsilon_{rd}^S$) that overlaps with the promoter. By enumerating the different states with their associated weights in a way similar to that exploited in fig.2a, and noting that the state where both the repressor and RNAP bind to their sites is not allowed, we can again derive the form for promoter occupation, eqn.5, but this time with the regulation factor,

$$F_{reg}(A) = \frac{1}{1 + \frac{R}{N_{NS}} e^{-\Delta\varepsilon_{rd}/k_B T}} \quad . \tag{9}$$

The above scheme can be further extended to describe co-regulation by two or more activators and/or repressors. For example, in the case of activation considered above, if the binding of the activator to its operator site is itself assisted by a helper protein, which might bind to an adjacent site [1], then the regulation factor still has the form given in eqn.8, but with the number of activators $A$ replaced by an effective number of activators,

$$A' = A \frac{1 + \frac{H}{N_{NS}} e^{-\Delta\varepsilon_{hd}/k_B T} e^{-\varepsilon_{ha}/k_B T}}{1 + \frac{H}{N_{NS}} e^{-\Delta\varepsilon_{hd}/k_B T}} \quad . \tag{10}$$

Note that the multiplicative factor in eqn.10 has the same form as eqn.8 except that now the number of helper molecules $H$ appears in the expression, and the interaction energy $\varepsilon_{ha}$ refers to that between the helper molecules and activators. In fact, this is the generic expression describing the recruitment of one DNA-binding protein by another, and is not limited to activator-RNAP recruitment. The introduction of the regulation factor allows for a discussion of various regulatory motifs in a unified way, as made explicit by Table 1. These examples will be discussed in the context of particular bacterial systems in the ensuing paper. The main point captured by this table is that the conceptual picture of thermodynamic models is identical regardless of regulatory motif and involves summing over all of the relevant states and culminates in the regulation factor, which as will be shown in paper 2 [1], is equal to the measurable fold-change of promoter activity.

As a final example, we consider the way in which DNA looping can play a role in dictating the regulation factor. Indeed, recent work by Vilar and Leibler [28**, 39] and others [40, 24] has shown how the thermodynamic models can be applied to regulatory control by looping. In the accompanying paper, we apply these ideas to the particular question of how such regulation depends upon the distance between the two binding sites and content ourselves here with a discussion of the conceptual basis. Two distinct looping scenarios are shown in fig. 3. In case (a), a repressor molecule, which can bind to two distinct regions on the DNA, loops out the intervening region. The classic example of this mode of action is the Lac repressor. In case (b), one protein (such as CRP) favorably bends the DNA so that a second activator can contact RNAP while paying a lower free energy cost than it would if it were acting alone. In both cases the free energy cost associated with making a DNA loop is outweighed by the benefit of additional binding energy between the repressor and DNA (case (a)) and between the activator and RNAP (case (b)).

In summary, the statistical mechanical framework described here can be used to consider a number of different regulatory motifs [11, 25, 27**, 29, 30, 41] as showcased in Table 1. The reader is reminded that in each of the cases considered in the table, the probability of promoter occupancy is given by eqn.5 with the sole change from one case to the next being the form adopted by the regulation factor itself.

## 3. Conclusion and Future Prospects

We argue that as a result of the increasingly quantitative character of data on gene expression, there is a corresponding need for predictive models. We have reviewed a series of general arguments about the way in which batteries of transcription factors work in generic ways to mediate transcriptional regulation. The models described here result in a number of important classes of predictions. The application of these ideas to particular bacterial scenarios forms the substance of the second article.

Though ideas like those presented here have the potential to serve as a quantitative framework for thinking about transcriptional regulation, there are a number of outstanding issues. Some especially troubling features of these models are: i) what are the precise conditions under which equilibrium assumptions are acceptable, ii) when can the probability of RNAP binding at a promoter serve as a surrogate for gene expression itself, iii) what is the role of fluctuations, iv) these models pretend that the basal transcription apparatus is a single molecule that interacts with transcription factors, whereas the transcription apparatus is a complex which is itself probably subject to recruitment for its assembly. Despite these concerns, our view is that these models have long demonstrated their utility and it will be of great interest to carefully explore their consequences experimentally. Case studies using the thermodynamic models is the mission of paper 2 [1].

## 4. Acknowledgements

We are grateful to a number of people for explaining their work and that of others to us, including Michael Welte, Jon Widom, Mark Ptashne, Phil Nelson, Jeff Gelles, Ann Hochschild, Mitch Lewis, Bob Schleif, Michael Elowitz, Paul Wiggins, Mandar Inamdar, Scott Fraser, Richard Ebright, Eric Davidson, Titus Brown. Of course, any errors in interpretation are our own. We gratefully acknowledge the support of the NIH Director's Pioneer Award (RP), NSF through grants DMR9984471 (JK) and DMR0403997(JK). JK is a Cottrell Scholar of Research Corporation. UG acknowledges an 'Emmy Noether' research grant from the *DFG*. TH is grateful to financial support by the NSF through grants 0211308, 0216576, and 0225630.

## References

[1] Bintu L, Buchler N, Garcia H, Gerland U, Hwa T, Kondev J, Kuhlman T, Phillips R: **Transcriptional Regulation by the Numbers 2: Applications**. *arXiv:q-bio.MN/0412011*
[2] ** Ptashne M: **A Genetic Switch**. Cold Spring Harbor Laboratory Press, Cold Spring Harbor, New York, 2004.
This book is a reprinting of Ptashne's classic with a special additional chapter that examines recent developments concerning regulation of the life cycle of phage lambda. One of the key recent developments is an appreciation of the role of DNA looping in this system.
[3] ** Ptashne M, Gann A: **Genes and Signals**. Cold Spring Harbor Laboratory Press, Cold Spring Harbor, New York, 2002.

This book gives an "extended argument" which illustrates in qualitative terms the notion of regulated recruitment and serves as the jumping off point for the class of thermodynamic model described here.

[4] Bellomy GR, Mossing MC, Record MT: **Physical Properties of DNA in Vivo As Probed by the Length Dependence of the *lac* Operator Looping Process**. *Biochem* 1988, **27**, 3900-3906.
[5] Oehler S., Amouyal M., Kolkhof P, vonWilcken-Bergmann B, Müller-Hill B: **Quality and position of the three *lac* operators of E. coli define efficiency of repression**. *EMBO J*. 1994, **13**, 3348-3355.
[6] Müller J, Oehler S, Müller-Hill B: **Repression of *lac* Promoter as a Function of Distance, Phase and Quality of an Auxiliary *lac* Operator**. *J. Mol. Biol.* 1996, 257, 21-29.
[7] Lee D-H, Schleif RF: ***In vivo* DNA loops in *araCBAD*: Size limits and helical repeat.** *Proc. Nat. Acad. Sci*. 1989, **86**, 476-480.
[8] Lewis DEA, Adhya S: **In Vitro Repression of the *gal* Promoters by GalR and HU Depends on the Proper Helical Phasing of the Two Operators**. *J. Biol. Chem.* 2002, **277**, 2498-2504.
[9] Hochschild A, Ptashne M: **Interaction at a distance between $\lambda$ repressors disrupts gene activation**. *Nature* 1988, **336**, 353-357.
[10] Joung JK, Koepp DM, Hochschild A: **Synergistic Activation of Transcription by Bacteriophage $\lambda$ cI Protein and *E. coli* cAMP Receptor Protein**. *Science* 1994, **265**, 1863-1866.
[11] Setty Y, Mayo AE, Surette MG, Alon U: **Detailed map of a cis-regulatory input function**. *Proc. Nat. Acad. Sci.* 2003, **100**, 7702-7707.
[12] Kalir S, McCluer J, Pabbaraju K., Southward C., Ronen M., Leibler S., Surette MG, Alon U.: **Ordering Genes in a Flagella Pathway by Analysis of Expression Kinetics from Living Bacteria**. *Science* 2001, **292**: 2080-2083.
[13] Laub MT, McAdams HH, Feldblyum T., Fraser CM, Shapiro L: **Global Analysis of the Genetic Network Controlling a Bacterial Cell Cycle**. *Science* 2000, **290**, 2144-2148.
[14] Arbeitman JN, Furlong EEM, Imam F, Johnson E., Null BH, Baker BS, Krasnow MA, Scott MP, Davis RW, White KP: **Gene Expression During the Life Cycle of *Drosophila melanogaster***. *Science* 2002, **297**, 2270-2275.
[15] ** Davidson EH: **Genomic Regulatory Systems**. Academic Press, San Diego, California, 2001.

Davidson's book provides an overview of common features of the cis-regulatory apparatus in different developmental settings.

[16] * Carroll SB, Grenier JK, Weatherbee SD: **From DNA to Diversity.** Blackwell Science, Malden, Massachusetts, 2001.

Like Davidson's book, Carroll et al. also argue for certain generic features of regulatory motifs that make generic statistical mechanics arguments more plausible as the basis for thinking about transcriptional regulation.

[17] Small S, Blair A, Levine M: **Regulation of *even-skipped* stripe 2 in the *Drosophila* embryo**. *EMBO J.* 1992, **11**, 4047-4057.
[18] Ackers GK, Johnson AD, Shea MA: **Quantitative model for gene regulation by $\lambda$ phage repressor**. *Proc. Nat. Acad. Sci.* 1982, **79**, 1129-1133.

[19] Shea MA, Ackers GK: **The OR Control System of Bacteriophage Lambda, A Physical-Chemical Model for Gene Regulation**. *J. Mol. Biol.* 1985, **181**, 211-230.
[20] Hill TL: **Cooperativity Theory in Biochemistry**. Springer-Verlag, New York, New York, 1985.
[21] Von Hippel PH, Revzin A, Gross CA, Wang AC**: Non-specific DNA Binding of Genome Regulating Proteins as a Biological Control Mechanism: 1. The** *lac* **Operon: Equilibrium Aspects**. *Proc. Nat. Acad. Sci.* 1974, **71**, 4808-4812.
[22] Law SM, Bellomy GR, Schlax PJ, Record MT: **In Vivo Thermodynamic Analysis of Repression with and without Looping in lac Constructs**. *J. Mol. Biol.* 1993, **230**, 161-173.
[23] Ben-Naim A: **Cooperativity in binding of proteins to DNA**. *J. Chem. Phys.* 1997, **107**, 10242-10252; **Cooperativity in binding of proteins to DNA. II. Binding of bacteriophage λ repressor to the left and right operators**, 1998, **108**, 6937-6946.
[24] Dodd IB, Shearwin KE, Perkins AJ, Burr T, Hochschild A, Egan JB: **Cooperativity in the long-range gene regulation by the λ cI repressor**. *Genes and Dev.* 2004, **18**, 344-354.
[25] Bakk A, Metzler R, Sneppen K: **Sensitivity of OR in Phage λ**. *Biophys. J.* 2004, **86**, 58-66.
[26] Bakk A, Metzler R: **In vivo non-specific binding of λ CI and Cro repressors is significant**. *FEBS Letters* 2004, **563** 66-68.
[27] ** Buchler NE, Gerland U, Hwa T: **On schemes of combinatorial transcription logic**. *Proc. Nat. Acad. Sci.* 2003, **100**, 5136-5141.
The *Supporting Text* of this paper shows how to implement models like those described here in statistical mechanics language and applies it to construct various logical states.
[28] ** Vilar JMG, Leibler S: **DNA Looping and Physical Constraints on Transcriptional Regulation**. *J. Mol. Biol.* 2003, **331**, 981-989.
This paper provides an explicit calculation of repression in the *lac* operon and shows that this results in a consistent definition of the looping free energy.
[29] Gerland U, Moroz JD, Hwa T: **Physical constraints and functional characteristics of transcription factor-DNA interaction**. *Proc. Nat. Acad. Sci.* 2002, 99, 12015-12020.
[30] Sengupta AM, Djordjevic M, Shraiman BI: **Specificity and robustness in transcription control networks.** *Proc. Nat. Acad. Sci.* 2002, 99, 2072-2077.
[31] Bruinsma RF: **Physics of Protein-DNA Interaction.** In *Physics of Bio-molecules and Cells*. Edited by Flyvbjerg H, Julicher F, Ormos P and David F. Springer-Verlag; 2002.
[32] Molineux IJ: **No syringes please, ejection of phage T7 DNA from the virion is enzyme driven**. *Mol. Microbiol.* 2001, **40**, 1-8.
[33] Record MT, Reznifoff WS, Craig ML, McQuade KL, Schlax PJ: *Escherichia coli* **RNA polymerase ($\sigma_{70}$) promoters and the kinetics of the steps of transcription initiation.** In *Eschericia coli and Salmonella Cellular and Molecular Biology**. Edited by Neidhardt FC et al. ASM Press, Washington DC; 1996: 792-821.
[34] Straney SB, Crothers DM: **Kinetics of the stages of transcription initiation at the** *Escherichia coli lac* **UV5 promoter**. *Biochem* 1987, **26**, 5063-5070.
[35] Strauss HS, Burgess RR, Record MT: **Binding of Escherichia coli ribonucleic acid polymerase to a bacteriophage T7 promoter-containing fragment: Evaluation of

**promoter binding constants as a function of solution conditions**. *Biochemistry* 1980, **19**, 3504-3515.

[36] Fried MG, Crothers DM: **Equilibrium studies of the cyclic-amp receptor protein-DNA interaction**. *J. Mol. Biol*. 1984, **172**, 241-262.

[37] Wong P, Gladney S, Keasling JD: **Mathematical model of the *lac* operon: Inducer exclusion, catabolite repression, and diauxic growth on glucose and lactose**. *Biotech. Prog*. 1997, **13**, 132-143.

[38] Cook DI, Revzin A: **Intracellular Localization of Catabolite Activator Protein of *Escherichia coli***. *J. Bacteriol.* 1980, **141**, 1279.

[39] Vilar JMG, Leibler S: paper in this volume.

[40] Seabold RR, Schleif RF: **Apo-AraC Actively Seeks to Loop**. *J. Mol. Biol*. 1998, **278**, 529.

[41] Aurell E, Brown S, Johanson J, Sneppen K: **Stability puzzles in phage $\lambda$**. *Phys. Rev*. 2002, **E65**, 051914-1-051914-9.

[42] Joung JK, Le LU, Hochschild A: **Synergistic activation of transcription by *Escherichia coli* cAMP receptor protein**. *Proc. Nat. Acad. Sci.* 1993, **90**, 3083.

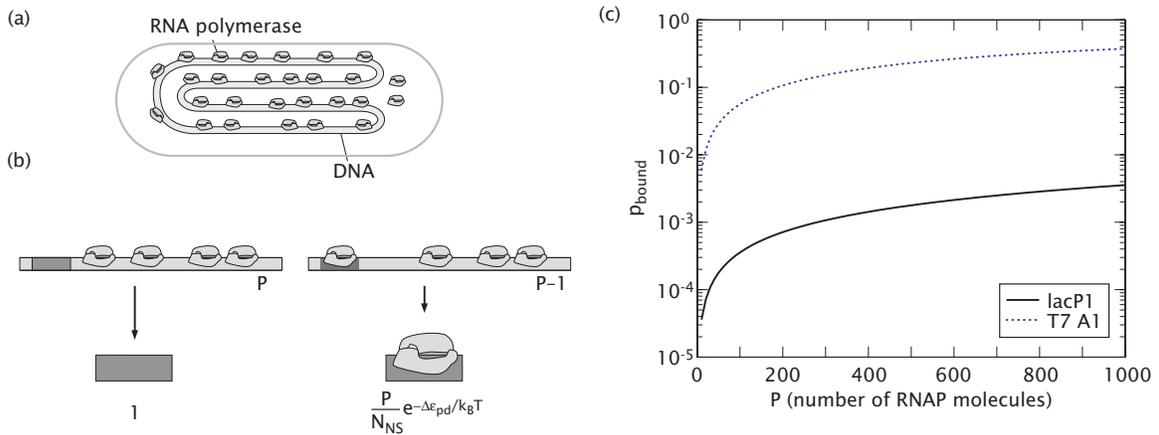

Figure 1: (a) Cartoon showing how, in the simple model, the DNA molecule serves as a reservoir for the RNAP molecules, almost all of which are bound to DNA. (b) Illustration of the states of the promoter - either with RNAP not bound or bound and the remaining polymerase molecules distributed among the nonspecific sites. The statistical weights associated with these different states of promoter occupancy are also shown. (c) Probability of binding of RNAP to promoter as a function of the number of RNAP molecules for two different promoters. We assume the number of nonspecific sites is $N_{NS} = 5 \times 10^6$ and compute the binding energy difference using the simple relation $\Delta \varepsilon_{pd} = k_B T \ln(K_{pd}^S / K_{pd}^{NS})$, where the equilibrium dissociation constants for specific ($K_{pd}^S$) and nonspecific binding ($K_{pd}^{NS}$) are taken from in vitro measurements. In particular, making the simplest assumption that the genomic background for RNAP is given also by the non-specific binding of RNAP with DNA, we take $K_{pd}^{NS} = 10,000$nM [33], for the *lac* promoter $K_{pd}^S = 50$nM [34] and for the T7 promoter, $K_{pd}^S = 1$nM [35]. For the *lac* promoter this results in $\Delta \varepsilon_{pd} = -5.3 k_B T$ and for the T7 promoter, $\Delta \varepsilon_{pd} = -9.21 k_B T$.

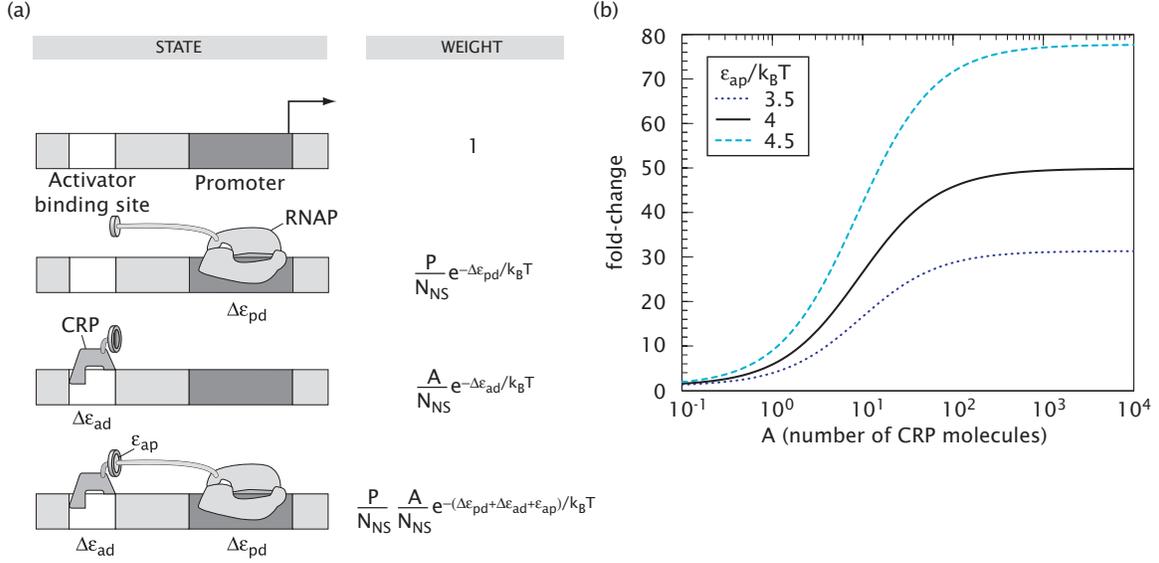

Figure 2: (a) Schematic showing the relation between the cartoon model of the various states of the promoter and its regulatory region, and their corresponding weights within the statistical mechanics framework. (b) Fold-change in promoter activity as a function of the number of activator molecules according to eqns. 5 and 8, for different values of the adhesive interaction energy between activator and RNAP. As in fig. 1,
$\Delta \varepsilon_{ad} = k_B T \ln(K_{ad}^S / K_{ad}^{NS})$
with $K_{ad}^{NS} = 10,000\text{nM}$ [36] and $K_{ad}^S = 0.02\text{nM}$ [37]. These *in vitro* numbers are chosen as a representative example to provide intuition for the action of activators - applications to *in vivo* experiments are given in paper 2 [1]. Several different representative values of the adhesive interaction $\varepsilon_{ad}$ that are consistent with measured activation are chosen to illustrate how activation depends upon this parameter.

| Case | Regulation Factor ($F_{reg}$) | |
|---|---|---|
| **1.** Simple repressor 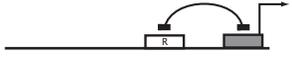 | $(1+r)^{-1}$ | $\left(1+\frac{[R]}{K_R}\right)^{-1}$ |
| **2.** Simple activator 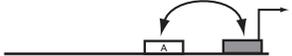 | $\dfrac{1+ae^{-\frac{\varepsilon_{ap}}{k_BT}}}{1+a}$ | $\dfrac{1+\frac{[A]}{K_A}f}{1+\frac{[A]}{K_A}}$ |
| **3.** Activator recruited by a helper (H) 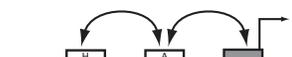 | $\dfrac{1+a\frac{1+he^{-\frac{\varepsilon_{ha}}{k_BT}}}{1+h}e^{-\frac{\varepsilon_{ap}}{k_BT}}}{1+a\frac{1+he^{-\frac{\varepsilon_{ha}}{k_BT}}}{1+h}}$ | $\dfrac{1+\frac{[H]}{K_H}+\frac{[A]}{K_A}f+\frac{[A]}{K_A}\frac{[H]}{K_H}f\omega}{1+\frac{[H]}{K_H}+\frac{[A]}{K_A}+\frac{[A]}{K_A}\frac{[H]}{K_H}\omega}$ |
| **4.** Repressor recruited by a helper (H) 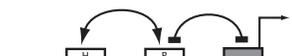 | $\left(1+\frac{1+he^{-\frac{\varepsilon_{hr}}{k_BT}}}{1+h}r\right)^{-1}$ | $\dfrac{1+\frac{[H]}{K_H}}{1+\frac{[H]}{K_H}+\frac{[R]}{K_R}+\frac{[R]}{K_R}\frac{[H]}{K_H}\omega}$ |
| **5.** Dual repressors 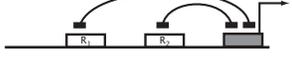 | $(1+r_1)^{-1}(1+r_2)^{-1}$ | $\left(1+\frac{[R_1]}{K_{R_1}}\right)^{-1}\left(1+\frac{[R_2]}{K_{R_2}}\right)^{-1}$ |
| **6.** Dual repressors interacting 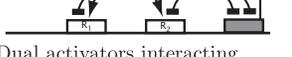 | $\left(1+r_1+r_2+r_1 r_2 e^{-\frac{\varepsilon_{r_1 r_2}}{k_B T}}\right)^{-1}$ | $\left(1+\frac{[R_1]}{K_{R_1}}+\frac{[R_2]}{K_{R_2}}+\frac{[R_1]}{K_{R_1}}\frac{[R_2]}{K_{R_2}}\omega\right)^{-1}$ |
| **7.** Dual activators interacting 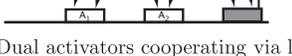 | $\dfrac{1+a_1 e^{-\frac{\varepsilon_{a_1 p}}{k_B T}}+a_2 e^{-\frac{\varepsilon_{a_2 p}}{k_B T}}+a_1 a_2 e^{-\frac{\varepsilon_{a_1 p}+\varepsilon_{a_2 p}+\varepsilon_{a_1 a_2}}{k_B T}}}{1+a_1+a_2+a_1 a_2 e^{-\frac{\varepsilon_{a_1 a_2}}{k_B T}}}$ | $\dfrac{1+\frac{[A_1]}{K_{A_1}}f_1+\frac{[A_2]}{K_{A_2}}f_2+\frac{[A_1]}{K_{A_1}}\frac{[A_2]}{K_{A_2}}f_1 f_2\omega}{1+\frac{[A_1]}{K_{A_1}}+\frac{[A_2]}{K_{A_2}}+\frac{[A_1]}{K_{A_1}}\frac{[A_2]}{K_{A_2}}\omega}$ |
| **8.** Dual activators cooperating via looping 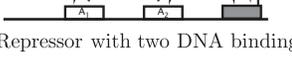 | $\dfrac{1+a_1 e^{-\frac{\varepsilon_{a_1 p}}{k_B T}}+a_2 e^{-\frac{\varepsilon_{a_2 p}}{k_B T}}+a_1 a_2 e^{-\frac{\varepsilon_{a_1 p}+\varepsilon_{a_2 p}+F_{loop}}{k_B T}}}{(1+a_1)(1+a_2)}$ | $\dfrac{1+\frac{[A_1]}{K_{A_1}}f_1+\frac{[A_2]}{K_{A_2}}f_2+\frac{[A_1]}{K_{A_1}}\frac{[A_2]}{K_{A_2}}f_1 f_2\omega}{(1+\frac{[A_2]}{K_{A_2}})(1+\frac{[A_1]}{K_{A_1}})}$ |
| **9.** Repressor with two DNA binding units and DNA looping 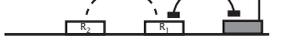 | $\left(1+r_1+\frac{r_1}{1+r_2}e^{-\frac{\Delta\varepsilon_{rd_2}+F_{loop}}{k_B T}}\right)^{-1}$ | $\dfrac{1+\frac{[R]}{K_2}}{\left(1+\frac{[R]}{K_1}\right)\left(1+\frac{[R]}{K_2}\right)+\frac{[R]\cdot[L]}{K_1\cdot K_2}}$ |
| **10.** N nonoverlapping activators and/or repressors acting independently on RNAp 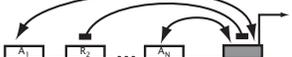 | $F_{reg1}\cdot F_{reg2}\cdot\ldots\cdot F_{regN}$ | $F_{reg1}\cdot F_{reg2}\cdot\ldots\cdot F_{regN}$ |

Table 1: Regulation factors for a number of different regulatory motifs. In the schematics of the motifs appearing in the first column, the symbol ⊥ indicates repression, ↓ represents activation, and a dashed lined is for DNA looping. The second column gives the regulation factor in terms of the number of transcription factors (TFs) in the cell and their binding energies, while the third column provides a translation of the regulation factor into the language of concentrations and equilibrium dissociation constants (used in paper 2 [1]). For an arbitrary TF we introduce the following notation: in the second column, $x$ is the combination $\dfrac{X}{N_{NS}}e^{-\Delta\varepsilon_{xd}/k_BT}$, while $[X]$ in the third column denotes the concentration of transcription factor X. $K_X = [X]/x$ is the effective equilibrium dissociation constant of the TF and its operator sequence on the DNA. Furthermore, in the third column we introduce $f = e^{-\varepsilon_{xp}/k_BT}$ for the "glue-like" interaction of a TF and

RNAP, and $\omega = e^{-\varepsilon_{x_1x_2}/k_BT}$ for the interaction between two TFs. In entries 8 and 10, $F_{loop}$ is the free energy of DNA looping, $\omega$ in 8 is defined as $e^{-F_{loop}/k_BT}$, while $[L]$ in 9 is the combination $\frac{N_{NS}}{V_{cell}} e^{-F_{loop}/k_BT}$, $V_{cell}$ being the volume of the cell.

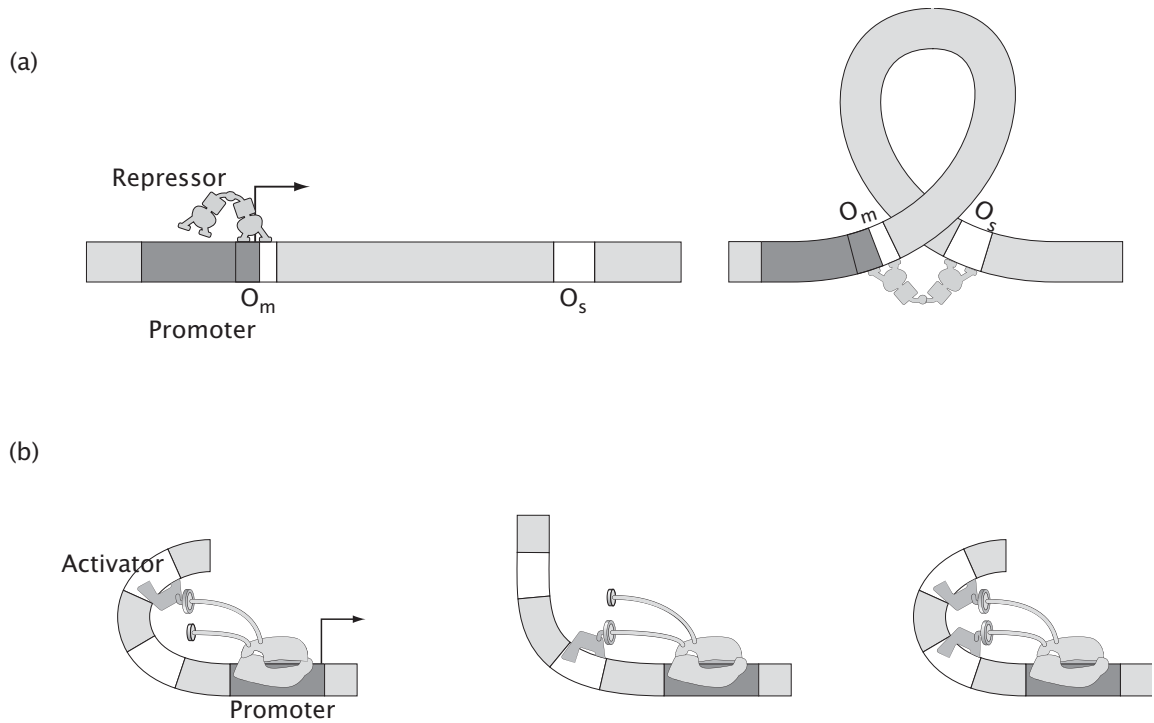

Figure 3: Schematic showing how DNA looping can be used both in repression and activation. (a) DNA looping allows Lac repressor to bind to the primary and the secondary operators simultaneously thereby increasing the weight of the states in which the promoter is unoccupied. This leads to stronger repression than in the single operator case. (b) DNA bending by the activator leads to cooperative binding of the two activators since the free energy cost of bending is paid only once. This leads to a boost in activation above that provided by independent binding of the two activators [42].